# Long-wavelength infrared photovoltaic heterodyne receivers using patch-antenna quantum cascade detectors

Azzurra Bigioli,[1,4] Giovanni Armaroli,[1] Angela Vasanelli,[1] Djamal Gacemi,[1] Yanko Todorov,[1] Daniele Palaferri,[1,3] Lianhe Li,[2] A. Giles Davies,[2] Edmund H. Linfield,[2] Carlo Sirtori [1,5]

[1] Laboratoire de Physique de l'Ecole Normale supérieure, ENS, Université PSL, CNRS, Sorbonne Université, Université de Paris, 24 rue Lhomond, 75005, Paris, France

[2] School of Electronic and Electrical Engineering, University of Leeds, LS2 9JT Leeds, UK

[3] Present address: GEM Elettronica srl, Via Amerigo Vespucci 9, 63074 San Benedetto del Tronto, Italy

[4]Corresponding author: azzurra.bigioli@ens.fr

[5]Corresponding author: carlo.sirtori@ens.fr

**Quantum cascade detectors (QCD) are unipolar infrared devices where the transport of the photo excited carriers takes place through confined electronic states, without an applied bias. In this photovoltaic mode, the detector's noise is not dominated by a dark shot noise process, therefore, performances are less degraded at high temperature with respect to photoconductive detectors. This work describes a 9 μm QCD embedded into a patch-antenna metamaterial which operates with state-of-the-art performances. The metamaterial gathers photons on a collection area, $A_{coll}$, much bigger than the geometrical area of the detector, improving the signal to noise ratio up to room temperature. The background-limited detectivity at 83 K is $5.5 \times 10^{10}$ cm Hz$^{1/2}$ W$^{-1}$, while at room temperature, the responsivity is 50 mA/W at 0 V bias. Patch antenna QCD is an ideal receiver for a heterodyne detection set-up, where a signal at a frequency 1.4 GHz and T=295 K is reported as first demonstration of uncooled 9μm photovoltaic receivers with GHz electrical bandwidth. These findings guide the research towards uncooled IR quantum limited detection.**

Highly sensitive photodetection in the long-wavelength infrared radiation range (LWIR), where photons have energies of the order of $\hbar\omega \sim$ 100-200 meV, is a challenging open problem essential to many sensing applications. In this spectral range, power detectors are hindered by thermally activated dark current, which binds their operation at cryogenic temperatures. A possible solution to this issue, proposed after the discovery of the $CO_2$ lasers, is the use of the amplification provided by beating, on a fast detector, the weak signal with a powerful local oscillator shifted in frequency. [1] In this configuration, known as heterodyne, the signal to noise ratio will be ultimately limited by the quantum efficiency of the detector, independently of the dark current. Single line gas-lasers are now replaced by quantum cascade lasers which offer ~100 mW of power and can be precisely frequency-tuned with temperature.[2,3] This is a great advantage for the heterodyne scheme which relies today on compact and efficient semiconductor local oscillators.

Quantum well infrared photodetectors (QWIP) [4] and quantum cascade detectors (QCD) [5] have two properties, related to the extremely short (~1ps) lifetime of intersubband transitions (ISBTs), that make these devices unique for heterodyne detection: the very high frequency response and saturation intensity. In the case of QWIPs, a signal up to a frequency of 82 GHz [6] and linear response under laser intensities in the kW/cm$^2$ range have already been demonstrated. [7]

A class of highly sensitive heterodyne receivers in the mid-infrared is required today for promoting technological applications and answering fundamental physical questions. This is relevant in observational astronomy [8] and high resolution spectroscopy, already in demand for the development of low-noise and high frequency detection systems. [9, 10] Ultrafast detectors are also required for coherent free-space LWIR communication platforms and light detection and ranging (LIDAR) systems.[11, 12] In perspective, quantum-limited heterodyne detection will also enable on-chip readout of photon statistics and quantum noise in the IR. In this work we have fabricated room temperature, photovoltaic heterodyne receivers at λ ~ 9μm by embedding a QCD into an antenna-resonator metamaterial. The device is a GaAs/AlGaAs QCD containing 8 periods, each composed of 5 quantum wells (QWs). It is designed to absorb at a wavelength of λ = 9 µm (140 meV). The schematized band diagram of the active region is shown in Fig. 1a. The green curves represent the square moduli of the wave functions involved in the photodetection and photoelectrons extraction process. The excited photoelectrons tunnel into the resonant level of the second well, and then relax toward the next active well by longitudinal optical phonon scattering through three intermediate levels (black lines) distant of approximatively $\hbar\omega_{LO} \sim 36$ meV. The first QW of each period is Si-doped $n_{2D} = 5.0 \times 10^{11}$ cm$^{-2}$. The structure has been inserted in an array of double-metal patch resonators, which provide sub-wavelength electric field confinement and act as antennas and contacts. [13,14,15] The top layer is a Pd/Ge/Ti/Au ohmic contact and serves to extract the photocurrent. The array is visible in the electron microscopy image of Fig. 1 b. Each array is composed of 15x15 patches, electrically connected by 130 nm wide wires where a Ti/Au Schottky contact has been evaporated.

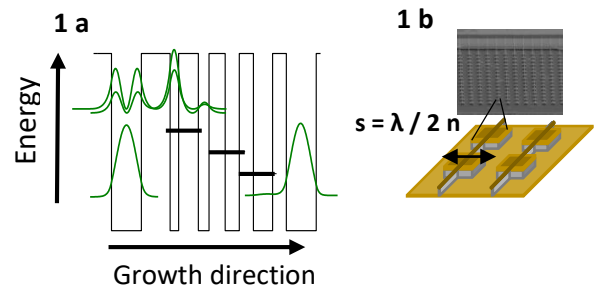

**Fig 1**. a) Band diagram of the QCD. The square moduli of the wavefunctions involved in the detection are indicated in green. The black lines represent the extraction levels. The layer thicknesses in nm are **7**/6.7/2/**4.6**/2.5/**3.8**/3.3/**3.3**/4.5/**3.2** with the first underlined layer doped at $n_{3D} = 5 \times 10^{11}$ cm$^{-3}$. Barriers are indicated in bold. b) SEM image of the metamaterial detector, sketched in the lower panel.

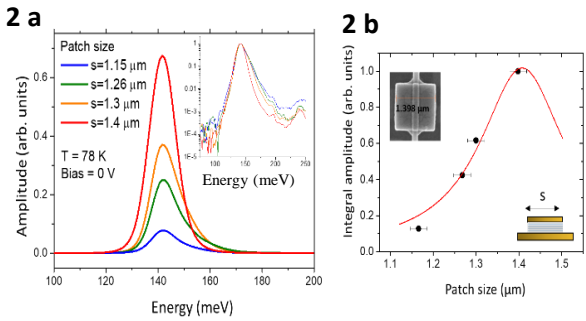

**Fig. 2**. a) Photocurrent spectra measured at 78 K and 0 bias for 4 different patch sizes. In the inset, the same measurements in a normalized log scale. b) Integral amplitude of the photocurrent spectra at varying patch size. The error bar of 20 nm is EBL design error. The red curve is a Lorentzian model accounting for the combined intersubband and cavity absorption (see text). In the upper-left corner, a SEM top-view of the 1.4 µm single patch; at the bottom right, a schematic lateral view of a single double metal

The distance between each patch is fixed at a= 2µm, where optimized photon absorption (coupling efficiency of 80%) has been confirmed by reflectivity measurements. [13] The resonant wavelength is defined by the lateral patch size s according to $\lambda = 2sn_{eff}$, where $n_{eff} = 3.2$ is the effective index. [15]

Patches of different dimensions have been processed in order to determine the structure with a mode resonant with the intersubband absorption, thus optimizing the cavity effect. Fig. 2a) illustrates photocurrent spectra for 4 devices with patch sizes s=1.15, 1.26, 1.3, 1.4 µm measured at 78 K and 0 V. In the inset of Fig. 2a) normalized photocurrent spectra are shown in a logarithmic scale to highlight the spectral broadening related to the shift of the cavity resonance towards higher energies for small cavity size. For s = 1.4 µm the spectrum has a symmetric shape as the two peaks, associated with the cavity mode and the intersubband transition, converge into resonance. The integral of the spectra is proportional to the responsivity of the device and is plotted in Fig. 2b as a function of the patch lateral size, s. The integral values show a net drop of the responsivity when the cavity is detuned from the intersubband transition. The x-error-bars are associated to a 20 nm offset of the electron beam lithography (EBL) patterning. The red curve, in Fig. 2b, is obtained by integrating the product of the measured spectral response of the bare detector, $S_{ISB}(E)$, times the microcavity absorptivity $S_{cavity}(E)$. $S_{ISB}(E)$ is the same in all processed samples, while $S_{cavity}(E)$ is modeled as a Lorentzian curve that peaks at different energies as a function of the size s. $S_{cavity}(E)$ has a quality factor Q = 4.7, extrapolated from reflectivity measurements on similar patch devices.[16] The model reproduces accurately the data and suggests that the optical cavity acts as a broad band-pass filter on the photocurrent of the bare intersubband transition. Notice that the spectra peak always at the same energy (Fig. 2a), as the linewidth of the bare photocurrent spectrum is narrower than that of the cavity. The model indicates optimal performances for s = 1.41 µm, confirming that the measured 1.4µm device is, within the error, resonant with the optical transition. In the rest of the article we concentrate on the performance characterization of the resonant device.

Patch antenna microcavities enable the coupling of normal incident light to ISB transitions and enhance the detector signal to noise ratio, as the antennas permit to gather photons on a collection-area $A_{coll}$, which is much bigger than its geometrical surface σ. The signal to noise ratio of the detector is expressed by the background limited detectivity $D^*_{BL}$, reported in figure 3a as a function of the temperature. Here, the values of the detectivity at 0 V for the patch-antenna detector, in red dots, are compared to those in blue of a device with the same active region but processed in a mesa geometry (i.e. without the patch array). At low temperatures the QCD noise is dominated by the background current noise. In this limit the advantage of the patch antenna device is given by the cavity effect that increases the effective interaction length of the light with the ISB transitions. This enhances the absorption rate and therefore the responsivity. The measured ratio $D^*_{BL,patch}/D^*_{BL,mesa}$ of 2.1 at low temperatures is consistent with the square root ratio of the responsivity for the patch and mesa device $\sqrt{R_{patch}/R_{mesa}} = 2.3$, in agreement with the description given in the reference [13]. At these temperatures the background-limited detectivity $D^*_{BL}$ (T) for the patch device is 5.5 x 10$^{10}$ cm Hz$^{1/2}$ W$^{-1}$, which sets the patch-antenna QCD at the ideal blackbody-limited detection at 9μm. [17] Notably, the reduction of the dark current in the patch-antenna array increases the BLIP (background limited infrared photodetector) temperature from 69 K to 82 K.

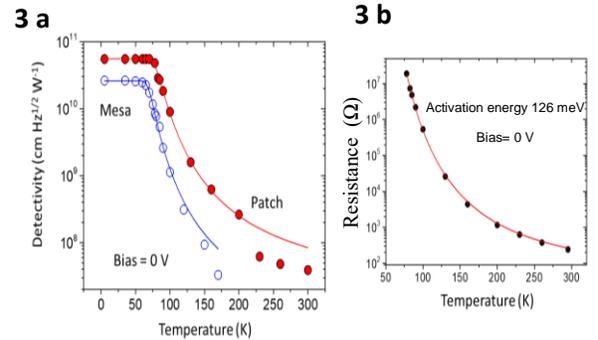

**Fig. 3**. a) Background limited detectivity as a function of the temperature for the mesa device as blue dots and for the patch detector in red. The fit curves are calculated using equation 1 and the resistance in figure b). The measured resistance of the device as a function of the temperature, in black dots, and the Schottky diode model, in red, using the activation energy as fitting parameter.

At high temperatures, the detector noise is dominated by the Johnson noise which is proportional to the electrical resistance of the device, Ω. The measured resistance, extracted from the dark current-voltage measurements, as a function of the temperature for the patch device is shown in figure 3b. In a patch antenna device, the electrical surface $\sigma$ is reduced with respect to its collection photon area $A_{coll}$ is reduced, thus increasing the device resistance of a factor $A_{coll}/\sigma$ This becomes apparent in the high temperature limit of the detectivity, $D^*_{BL}$, which can be written as:

$$1) \quad D^*_{BL}(T) = \frac{R(T)}{\sqrt{\frac{4kT}{\Omega_{surf}}}}\sqrt{\frac{A_{coll}}{\sigma}},$$

where $\Omega(T) = \Omega_{surf}(T)\sigma$ is a surface resistance in unit of [ohm cm²] and R is the responsivity. This expression of the detectivity, $D^*_{BL}$, is significant as it underlines the net improvement on the signal to noise that can be obtained at high temperatures in our photodetection process by increasing $A_{coll}$ with respect to $\sigma$.

Notice that $D^*_{BL,patch}$ at 300 K has the same value than $D^*_{BL,mesa}$ at 160 K, which means an increase of 140 K of the operating temperature in agreement with the previous results for quantum well infrared photodetectors having the same ratio $A_{coll}/\sigma$. [13] The temperature dependence of the resistance of the patch-antenna quantum cascade detector can be interpreted in analogy with a Schottky diode model.[18] In this approximation carrier transport is described as a diffusion current from the doped well (metal) to the cascade region (depletion region in the semiconductor) in a low mobility regime. The resistance values in Fig. 3b are therefore fitted with a function proportional to $(k_BT)^2 e^{E_A/k_BT}$ where the activation energy $E_A$ (analog to the barrier in a Schottky diode) is left as an adjustable parameter. The fit reproduces the data accurately for $E_A = 126$ meV across all the temperature range. This value of $E_A$ is also in excellent agreement with the activation energy that was measured from the thermally activated current and corresponds to the energy difference between the excited state of the optical transition and the Fermi energy (124 meV) obtained from simulations. By inserting this expression for the resistance $\Omega(T)$ in equation 1, it is possible to model the detectivity values as a function of the temperature, as shown in Fig. 3a as blue line for the mesa structure and red line for the patch-antenna device, considering the ratio $A_{coll}/\sigma$. Data are exactly described until the temperature of 200 K. The loss of accuracy in the detectivity model at high temperatures may stem from thermally induced charge depletion of the active well that causes an internal electric field. This may reduce the tunneling extraction of electrons from the excited state of the optical transition with a negative impact on the responsivity. However, the tunneling alignment can be adjusted by applying a voltage to the device, as it can be seen in the data reported in Fig. 4 showing responsivity measurements and photocurrent spectra as a function of the voltage at room temperature.

From Fig. 4a it can be observed that the responsivity exhibits a record value of 50 mA/W at 0 V bias. Previously, QCD responsivity of 16.9 mA/W at 300K and λ ∼ 9 μm, was achieved with an optical diagonal transition design. [19] This value can be increased for negative voltages, reaching 77 mA/W at -0.3 V, while for positive voltages the detector's response rapidly drops to 0. The same trend is confirmed by photocurrent spectra at different voltages, presented in Fig. 4b.

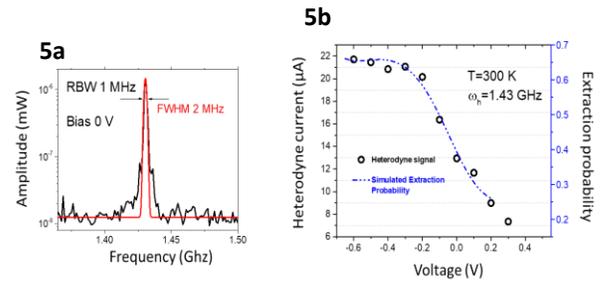

**Fig. 5** a) Heterodyne signal measured by the spectrum analyzer with a RBW = 1 MHz and no applied bias. The heterodyned optical pump beam has a Gaussian spectrum (red curve) with FWHM of 2 MHz at -3dBm b) In the right axis, heterodyne current (black dots), as a function of the bias. In the left axis, the extraction probability (black dashed line).

We demonstrate that these devices operate as heterodyne receivers at room temperature and at zero bias, in the GHz range. Our setup is made of two distributed feedback (DFB) quantum cascade lasers passively stabilized using low-noise voltage supplier and an optical isolator to prevent optical-feedback. Figure 5a presents a heterodyne signal at room temperature obtained with a local oscillator providing $P_{LO}$ = 4 mW. The signal $P_S$ is measured with a resolution bandwidth (RBW) of 1 MHz at a frequency of $\omega_h$ = 1.4 GHz and with no applied bias. The observed signal has a linewidth of ~2 MHz, which is dominated by the linewidth of the two free-running DFB lasers. [20]. The heterodyne current, $I_{het}$, as a function of the bias is shown in figure 5b (black dots). $I_{het}$ is proportional to the responsivity $R(V)$ defined in equation through the formula $I_{het} = 2R(V)\sqrt{P_{LO}P_s}cos(\omega_h t)$ where $P_{LO}$ is the local oscillator power and $P_s$ is the signal power. We remind that the intrinsic responsivity $R(V)$ of the QCD can be modeled as: [21]

$$2) \quad R(V) = \frac{\lambda}{h}\frac{q}{c}\frac{p_e(V)}{N_p}\eta$$

Where $\eta$ is the absorptivity, which depends on doping density, and $p_e$ is the extraction probability per period, defined as $p_e = \tau_{2,3\rightarrow 1}/(\tau_{escape} + \tau_{2,3\rightarrow 1})$ Here $1/\tau_{2,3\rightarrow 1}$ is the relaxation rate back to the ground state of the main well from the levels 2 and 3, as schematized in the inset in Fig. 4a, and $1/\tau_{escape}$ is the escape rate for reaching the next cascade structure and generating photocurrent. While the absorptivity can be taken constant with the voltage, the escape rate strongly depends on the applied bias that controls the strength of the tunnel couplings.

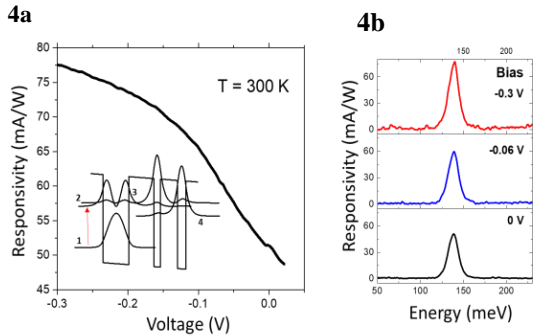

**Fig 4.** a) Responsivity curve as a function of the bias at room temperature. In the inset, the band diagram under applied positive voltage. b) Responsivity spectra for biases 0V, -0.06V

The effect of the bias on photoresponse is compared (blue dashed line in figure 5b) with the extraction probability, which is calculated within a model considering resonant tunneling transport of the electrons through the barrier.[22] The incoherent mechanisms in the transport are also included via longitudinal optical phonon and roughness scattering rates. [23] Referring to the band diagram in the inset of fig. 4 a, the escape rate can be evaluated using the following expression:[24, 25]

$$3) \quad \frac{1}{\tau_{escape}}(V) = \frac{2|\Delta|^2 \tau_{deph}}{1+(E_{2-3}(V)/\hbar)^2 \tau_{deph}^2 + 4\Delta^2 \tau_3 \tau_{deph}}$$

where $\omega = 2\frac{\Delta}{\hbar}$ is the Rabi oscillation frequency, corresponding to the resonant tunneling process between subbands 2 and 3 with coupling energy $\Delta$, $E_{2-3}$ is the energy detuning from resonance that depends on bias. The intersubband lifetime $\tau_3$ is limited by the emission of optical phonons. The dephasing time is set $\tau_{deph} = 0.1$ ps, resulting in a broadening comparable to that of the linewidth (FWHM = 13 meV) of a mesa processed device.[26] This value is also in line with the estimate from our model. For a negative bias = -0.2 V, levels 2-3 start to align. In this resonant condition the tunneling is very efficient and the escape rate is solely limited by the inelastic scattering time $\tau_3 \sim 0.3$ ps.

**Funding** Qombs Project (grant agreement number 820419). hoUDINi Project (ANR-16-CE24-0020)

**Acknowledgments** The authors acknowledge S. Suffit and P. Filloux for clean room technical support and Prof. J. Faist and F. Kapsalidis (ETH, Zurich) for providing the DFB QCLs.

**Disclosures** The authors declare no conflicts of interest.

The data that support the findings of this study are available from the corresponding authors upon reasonable request.

When the applied bias is positive, levels 2-3 are rapidly detuned from resonance and electrons are more likely relaxing back to the main well, preventing the photocurrent. The extraction probability model describes the photocurrent behavior as a function of the applied bias. Ultimately, an ultra-sensitive heterodyne detection set-up will benefit of a detector designed to provide maximum signal at zero bias, where the detector noise is the lowest. Our result clearly indicates that patch antenna QCD are a viable solution to realize shot noise limited heterodyne receivers for mid-infrared detection at room temperature. As these devices operate at zero bias, they are not dominated by dark current and could become shot noise limited with a moderate LO power. In conclusion, we presented an AlGaAs/GaAs quantum cascade detector working at 9 μm embedded in a patch-antenna metamaterial, operating also as fast heterodyne receiver. The benefits of the enhanced photon collection result in a background detectivity at low temperature of 5.5 x $10^{10}$ cm $Hz^{1/2}$ $W^{-1}$, at the ideal photovoltaic blackbody-limit, and a record value responsivity of 50 mA/W at room temperature at zero bias. Promising results of fast photovoltaic heterodyne signals at room temperature are presented, which, with further technical developments, could pave the way toward mid-infrared uncooled few-photons power detection.